\documentstyle[aps,preprint]{revtex}
\begin{document}
\title{Softening and Broadening of the Zone Boundary Magnons in 
Pr$_{0.63}$Sr$_{0.37}$MnO$_3$}
\author{H. Y. Hwang$^{1}$, P. Dai$^2$, S-W. Cheong$^1$, G. Aeppli$^3$, 
D. A. Tennant$^2$, and H. A. Mook$^2$}
\address{$^1$Bell Laboratories, Lucent Technologies, Murray Hill, 
New Jersey 07974\\
$^2$Solid State Division, Oak Ridge National Laboratory, Oak Ridge,
Tennessee 37831\\ $^3$NEC Research Institute, 4 Independence Way,
Princeton, New Jersey 08540}
\maketitle
\begin{abstract}
We have studied the spin dynamics in Pr$_{0.63}$Sr$_{0.37}$MnO$_3$
above and below the Curie temperature $T_C=301$ K.  Three distinct new
features have been observed: a softening of the magnon dispersion at
the zone boundary for $T<T_C$, significant broadening of the zone
boundary magnons as $T\to T_C$, and no evidence for residual spin-wave
like excitations just above $T_C$.  The results are inconsistent
with double exchange models that have been successfully applied to
higher $T_C$ samples, indicating an evolution of the spin system with
decreasing $T_C$.
\end{abstract}
\pacs{PACS numbers: }

 
The revival in the study of manganites has led to a reexamination of
the unique coupling between magnetism and charge transport in these
materials.  We focus on perovskite manganites with a transition from a
high temperature paramagnetic insulator to a low temperature
ferromagnetic metal at the Curie temperature $T_C$.  The samples that
exhibit this behavior have been partially hole doped away from a
parent antiferromagnetic insulator, such as LaMnO$_3$, by divalent
substitution on the La site, such as La$_{0.7}$Ca$_{0.3}$MnO$_3$
\cite{r1}.  The Mn $3d$ levels, split by the oxygen octahedral crystal
field to a lower energy $t_{2g}$ triplet and a higher energy $e_g$
doublet, are filled according to Hund's rule such that all spins are
aligned on a given site by a large intra-atomic exchange $J_H$.
Electronic conduction arises from the hopping of an electron from
Mn$^{3+}$ to Mn$^{4+}$ with electron transfer energy $t$.  This
results in the ferromagnetic double exchange interaction between
localized S=3/2 spins (the core $t_{2g}$ triplet) mediated by the
hopping $e_g$ electron \cite{r2,rr2}.

Recently, it has been shown that although the highest $T_C$ materials
($T_C>325$ K) are reasonably described by the double exchange model,
with decreasing $T_C$ (and reduced electronic bandwidth), the dramatic
magnetotransport properties and increasingly first-order transition
require the incorporation of a strong Jahn-Teller based phonon
coupling \cite{r3,ttt,tt}.  This has been corroborated by a number of
studies showing the growing importance of static and dynamic lattice
distortions as the zero temperature insulating state is approached
\cite{pd,pr}.  Thus far, most studies of the spin dynamics have
focussed on high $T_C$ samples which have been shown to be consistent
with the double exchange model \cite{n,nnn,se}.  With decreasing $T_C$
however, important deviations may occur, as indicated by the
development of a prominent diffusive central peak near $T_C$ in
La$_{0.67}$Ca$_{0.33}$MnO$_3$ ($T_C$ = 250 K) \cite{nn}.  By studying
the spin dynamics in a sample with reduced $T_C$, we can test if the
spin system remains well described by simple double exchange (allowing
for renormalizations of $t$ and $J_H$), or just as for the charge
dynamics, the lattice coupling must be explicitly considered.
 
In this paper we report a neutron scattering study of the magnetic
dynamics in Pr$_{0.63}$Sr$_{0.37}$MnO$_3$ above and below $T_C=301$ K,
focussing on the zone boundary magnons.  At 10 K (0.03$T_C$), the
magnons are well-defined for all wave-vectors $q$ and exhibit a
significant deviation from the dispersion associated with the simplest
local moment description.  At 265 K (0.9$T_C$), the dispersion
relation has uniformly softened but maintains its $q$ dependence, and
significant broadening of the short wavelength magnons is observed.
At 315 K (1.05$T_C$), just 14 K above $T_C$, there is no evidence for
an inelastic magnetic peak at any $q$.  We argue that these
observations are at odds with double exchange calculations of the spin
excitations (including leading order self-energy corrections),
indicating qualitatively different spin dynamics in lower $T_C$
samples.
 
We used the floating zone method to grow a large single crystal of
Pr$_{0.63}$Sr$_{0.37}$MnO$_3$ with a mosaic spread of 0.5$^{\circ}$
full-width at half-maximum (FWHM).  The neutron scattering
measurements were carried out on the HB1 triple-axis spectrometer at
the High-Flux Isotope Reactor, Oak Ridge National Laboratory.  The
collimations were, proceeding from the reactor to the detector,
50'-40'-S-40'-70', and the final neutron energy was fixed at
$E_f$=13.5 or 30.5 meV.  The monochromator, analyzers and filters were
all pyrolytic graphite.  The sample was slightly orthorhombic and
twinned at the measured temperatures.  Even so, we assumed a cubic
lattice ($a_o=3.86$ \AA) because we could not resolve the effects of
the twinning within the spectrometer resolution.
 
Figure 1 shows the magnon dispersion along the [0,0,1], [1,1,0], and
[1,1,1] directions at 10 K, and along [0,0,1] at 265 K (0.9$T_C$).
Figure 2 shows some of the constant q scans from which Figure 1 was
obtained.  The experimentally observed magnon peaks were accurately
described by Gaussian fits.  To verify that our results are not
instrumental artifacts, we performed a 4-dimensional Monte Carlo
convolution of the experimental resolution function \cite{cn} and the
theoretical neutron scattering cross-section for damped spin waves.
We verified that the peak positions were relatively unchanged,
especially near the zone boundary saddle point, where significant
resolution-induced shifts could occur.  Our calculations place an
upper bound of 0.5 meV for the shift in the peak position due to
resolution effects.  Another possible source of misinterpretation is
that the broad peak at (0,0,1.4) at 265 K is actually a composite of
phonon as well as magnon peaks.  This was ruled out by measuring the
same point in the next zone, which showed the same lineshape, down in
intensity by the magnetic form factor.
 
Focussing first on the low temperature dispersion, a new feature we
have observed is the significant softening at the zone boundary, seen
in all directions.  The Heisenberg spin Hamiltonian, $H=-\sum_{ij}
J_{ij}{\bf S_i} \cdot {\bf S_j}$, couples the spins at site $R_i$ and
$R_j$ by $J_{ij}$.  In the linear approximation, the spin wave
dispersion relation is given by $\hbar\omega({\bf q})=\Delta+2S(J({\bf
0})-J({\bf q}))$, where $J({\bf q})=\sum_j J_{ij}\exp[i{\bf
q}\cdot(R_i-R_j)]$.  $\Delta$ allows for small anisotropies.  The
solid line in Figure 1 is the outcome of a fit for only nearest
neighbor interactions for $\xi<0.2$ (the cubic zone boundary occurs at
$\xi=0.5$), resulting in $\Delta=1.3\pm0.3$ meV and $2SJ_1=8.2\pm0.5$
meV.  Although this describes the data for $\xi<0.2$, the zone
boundary magnons are missed by 15-30 meV.  By comparison, the magnon
dispersion in La$_{0.7}$Pb$_{0.3}$MnO$_3$ ($T_C=355$ K) was found to
be well described by only nearest neighbor interactions of similar
strength \cite{nnn}.
 
The dashed lines in Figure 1 are a fit to the full data set including
up to fourth neighbor interactions, resulting in $\Delta=0.2\pm0.3$
meV, $2SJ_1=5.58\pm0.07$ meV, $2SJ_2=-0.36\pm0.04$ meV,
$2SJ_3=0.36\pm0.04$ meV, and $2SJ_4=1.48\pm0.10$ meV.  This accurately
follows the dispersion except near the [0,0,1] zone boundary, which
requires the next Fourier term in this direction, corresponding to
$J_8$.  Nevertheless, the fourth nearest neighbor fit quantifies the
remarkable result that {\it additional extremely long range
ferromagnetic couplings are required}.  Although $J_2$ and $J_3$ were
necessary to fit the data, the more important correction to nearest
neighbor coupling is $J_4$.  The long range and non-monotonic behavior
of $J({\bf q})$ required by the data seems to rule out a simple
Heisenberg Hamiltonian.
 
On warming, the dispersion relation uniformly softens, as can be seen
in Figures 1 and 2 for the [0,0,1] branch at 265 K.  At $\xi=0.14$,
near the zone center, the magnon peak shows no substantial changes
other than decreasing to lower energy as temperature is increased.  In
addition, it is resolution-limited at both 10 K and 265 K.  Just above
$T_C$, at 315 K, there is no evidence for the magnon peak as expected
for the long-wavelength excitations.  As the Brillouin zone is
traversed (Figure 3), magnon lifetime effects become apparent.  In
particular, on approaching the zone boundary at $\xi=0.5$, the 10 K
linewidth is substantially larger than the experimental resolution
width.  On warming to 265 K, the deconvolved widths, which we quote as
full-widths at half-maximum throughout this paper and its figures, are
nearly doubled, from their 10 K value of 8.4$\pm$0.5 meV to
13.2$\pm$1.9 meV.  Again, at 315 K$ >T_C$, no obvious magnon peak
remains.
 
An important issue to address is the role of the magnetic Pr ions,
which display the crystal electric field (CEF) level shown in Figure
2.  A recent powder neutron diffraction study of a manganite with
similar Pr concentration observed a refined ferromagnetic moment of
$\sim0.5 \mu_{B}$ at low temperature \cite{r5}.  Thus the proper
description of the total spin system involves a two component
non-Bravais lattice.  A bound on possible Pr-Mn interactions can be
placed by examining the crossing of the CEF level and the magnon
dispersion, for example near $\xi=0.2$ in the [0,0,1] branch.  Even
for weak coupling, this region is susceptible to the effects of mixing
due to energy degeneracy.  Experimentally, we find that the CEF level
varies smoothly through the crossing of the magnon branch without
dispersion.  The integrated intensities of both the CEF excitation and
the magnons show no anomaly at their crossing, and in particular, the
observed magnon linewidth presented in Figure 3 shows no feature at
the crossing. Also, the magnon dispersion passes smoothly through the
CEF level just as in La$_{0.7}$Pb$_{0.3}$MnO$_3$, which does not
contain a magnetic rare earth ion (Figure 1).  Thus, our experiments
show no evidence for Mn-Pr coupling with an effect on the Mn spin
dynamics.
 
To place our study of Pr$_{0.63}$Sr$_{0.37}$MnO$_3$ in proper
context, it is useful to compare with more conventional ferromagnets.
Perhaps the most relevant comparison is to the europium chalcogenides,
insulating ferromagnets that are considered ideal Heisenberg spin
systems with extended exchange interactions (next nearest neighbor
interactions are significant) \cite{r9,r10,hm}.  In EuO, both $J_1$
and $J_2$ are ferromagnetic, whereas in EuS, $J_2$ is a competing
antiferromagnetic interaction with $J_1$ (higher order terms are
negligible).  Although similar to Pr$_{0.63}$Sr$_{0.37}$MnO$_3$ in
this respect, the abrupt zone boundary softening we have observed (not
seen in the europium chalcogenides) requires a large $J_4$ while $J_2$
and $J_3$ are relatively small.  Despite the large difference in $T_C$
for EuO ($T_C=69.2$ K) and EuS ($T_C=16.6$ K), the low frequency spin
stiffness, magnon bandwidth, and $T_C$ are self-consistently related
via simple mean field theory.  Comparison of the two manganite samples
with $T_C = 355$ K and $T_C = 301$ K (Figure 1) shows that at low
energy, the magnon dispersions are almost identical, while the zone
boundary softening of the lower $T_C$ sample results in a $\sim69\%$
decrease in the magnon bandwidth from $\sim 108$ meV to $\sim 74$ meV.
Although La$_{0.7}$Pb$_{0.3}$MnO$_3$ is roughly consistent with mean
field theory, with decreasing $T_C$ this breaks down because the spin
stiffness does not diminish - the additional exchange constants have
been exactly balanced by a reduction in the effective $J_1$.
 
The spin dynamics of both EuO and EuS have been extensively studied
\cite{r11,r12}, and the damping near $T_C$ in both cases were observed
to vary consistent with predominant magnon-magnon damping
\cite{swt1,swt2}, with the linewidth $\Gamma$ varying as
$\Gamma\propto q^4\ln^2(k_BT/\hbar\omega_{\bf q})$ for
$\hbar\omega_{\bf q}\ll k_BT$, and $\Gamma\propto q^3$ for
$\hbar\omega_{\bf q}\gg k_BT$.  Although theoretically valid only in
limiting extremes, and for $(a_oq)^2\ll1$, $\Gamma(q)$ was observed to
rise smoothly across $\hbar\omega_{\bf q}= k_BT$ in EuS.  By contrast,
we have observed the lack of a $q$ dependence near the zone boundary
for $\xi>0.35$.  At present, it remains unclear whether our results
indicate another important damping channel (quasiparticles, lattice
distortions), or whether a revised spin wave theory which takes
account of the unusual spin wave dispersion for
Pr$_{0.63}$Sr$_{0.37}$MnO$_3$ is sufficient.  Finally, in both EuO and
EuS, at short wavelengths, a magnon-like peak was observed far above
$T_C$, whereas we did not observe such a feature just above $T_C$ in
Pr$_{0.63}$Sr$_{0.37}$MnO$_3$ for any $\xi$ in the [0,0,1] branch.
 
We now discuss possible explanations of our results.  Within a Stoner
model, the conduction band in the ferromagnetic state is exchange
split into a majority and minority band.  The magnon dispersion enters
the Stoner continuum at finite q and $\omega$, where quasiparticle
damping of spin waves occurs.  The ferromagnetic ground state of the
manganite, however, is rather unique in that there is complete
separation of the majority and minority band by a large $J_H$.  Thus
at low temperatures, the entire spin wave dispersion probably lies
below the triplet electron-hole pair excitation continuum.  At
temperatures approaching $T_C$, however, the ordered moment is
decreased and the carriers are no longer fully polarized, bringing
triplet electron-hole pair excitations to energies within the spin
wave band.  These excitations can be the decay products of low energy
magnons, and as their probability rises with temperature, give rise to
temperature-dependent magnon lifetimes.
 
The damping mechanism described above can be encapsulated in the
imaginary part of the magnon self-energy.  The real part contains the
magnon dispersion, including its deviations from a simple cosine form.
The magnon dispersion incorporating the lowest order self-energy
correction has been calculated in a Kondo lattice model Hamiltonian
with ferromagnetic spin-electron exchange \cite{se}.  For
$J_H/t=\infty$, the result was a simple cosine dispersion consistent
with a nearest-neighbor ferromagnetic Heisenberg exchange.  For finite
values of $J_H/t$, there are deviations from the simple result, but
the deviations do not fit our experimental observations.  In
particular, the reduction of the low frequency spin stiffness is more
significant than the reduction of the total bandwidth, contrary to the
experimental results emphasized earlier.  Beyond these
inconsistencies, however, the most important point is that with
decreasing $T_C$, $t$ is reduced.  $J_H$ is presumably unchanged,
being an intra-atomic energy.  Thus with decreasing $T_C$, corrections
of order $t/J_H$ should decrease, not increase.  Therefore, lowest
order perturbative corrections in $t/J_H$ do not capture our
experimental results.
 
The Kondo-lattice calculations discussed above have two limitations:
the rate of convergence is not well understood, and the effects of
varying the hole concentration $x$ are not explored.  These issues
have been studied in one dimension by exact diagonalization of double
exchange coupled spin rings for large $J_H$ \cite{r6,r7}.  Deviations
from a simple cosine dispersion are predicted, particularly for
extremely large or small doping ($x<0.2$ and $x>0.7$).  For
intermediate carrier concentrations (as studied here), however, a
cosine band is recovered.  For the two samples compared in Figure 1,
the nominal carrier concentration is not significantly different,
leading to the conclusion that a degree of freedom beyond $x$, $t$, or
$J_H$ is needed to account for our data.
   
One much discussed parameter beyond the purely electronic parameters
is the electron-lattice coupling, which is worth considering because
of the large and strongly temperature-dependent mean square
displacements $\langle {\bf u}^2 \rangle$ of Mn and O near $T_C$
\cite{pd,pr}.  The bare overlap integral between the Mn spins is
highly sensitive to the geometric arrangement of the Mn-O-Mn linkage,
particularly due to the directional nature of the Mn $d$ and O $p$
orbitals.  The softening of the well-defined zone boundary magnons at
low temperature would then be due to slow (on the scale of the zone
boundary spin wave energies) fluctuations towards doubling of the
nuclear unit cell.  That fluctuations of this type might exist is
clear from the myriad of lattice and charge-ordering instabilities in
the phase diagrams of the manganites.  Coupling to the lattice can
also contribute to the magnon lifetime effects in the manganites. In
particular, the rapid rise in $\langle {\bf u}^2\rangle$ with
temperature could give rise to a broadening distribution of effective
exchange constants, leading to the pronounced broadening of the short
wavelength magnons near $T_C$. Long wavelength magnons, of course,
would not be affected as the variations in $J$ would be averaged out.
This scenario is qualitatively consistent with the $T_C$ dependence of
the spin dynamics, in that the highest $T_C$ samples do not show the
strong feature in $\langle {\bf u}^2 \rangle$ \cite{n}.

We thank B. Batlogg, J. A. Fernandez-Baca, N. Furukawa, Y. B. Kim, P.
B. Littlewood, J. W. Lynn, P. Majumdar, A. Millis, B. I. Shraiman, S.
H. Simon, and C. M. Varma for useful discussions.  HYH and GA would
like to thank the HFIR staff at ORNL for their hospitality and
assistance.  The work at ORNL was supported by the U.S. DOE under
contract No.  DE-AC05-96OR22464 with Lockheed Martin Energy Research,
Inc.

\begin{figure} 
\caption{Magnon dispersions for $(0,0,\xi)$,
$(\xi,\xi,0)$, and $(\xi,\xi,\xi)$ (where $\xi=0.5$ is the cubic zone
boundary) at $T=10$ K and 265 K. The solid line is a fit to a
nearest-neighbor Hamiltonian for $T=10$ K and $\xi < 0.2$.  The dashed
line is a fit for all data including up to fourth nearest neighbors at
$T=10$ K.  The dotted line is the corresponding four neighbor fit for
$T=265$ K.  Also shown in squares are data for
La$_{0.7}$Pb$_{0.3}$MnO$_3$ at 10 K (from Ref. 10).}
\end{figure}
\begin{figure} 
\caption{Constant-$q$ scans at two different wave
vectors with the left panel close to the zone center and the right
panel close to the zone boundary. There is a dispersionless crystal
electric field (CEF) level at $\sim$12 meV from Pr (shown as dash-dot
line).  The solid circles are a constant-$q$ scan of the CEF level at
(0,0,1), the zone center.  At 265 K, the intensity of the CEF level
drops to undetectable levels and has therefore been ignored in the
data analysis at this temperature.  The $T=315$ K data has been fitted
with a simple Lorentzian line-shape (including a Gaussian). The other
data have been fitted to a damped harmonic oscillator, including 4D
convolution of the instrumental resolution as described in the text.
The dashed line is the instrumental response to spin-waves with
infinite lifetimes and the dispersion shown in Figure 1.}
\end{figure} 
\begin{figure}
\caption{Magnon line widths in the [0,0,1] direction extracted from
Gaussian fits with a sloping background. The solid and dashed lines
are guides to the eye.} 
\end{figure} 
\end{document}